\documentclass{article}

\usepackage{arxiv}

\usepackage[utf8]{inputenc} 
\usepackage[T1]{fontenc}    
\usepackage{hyperref}       
\usepackage{url}            
\usepackage{booktabs}       
\usepackage{amsfonts}       
\usepackage{nicefrac}       
\usepackage{microtype}      
\usepackage{lipsum}
\usepackage{graphicx}
\graphicspath{ {./images/} }
\usepackage{lineno}

\usepackage{amsmath, amssymb}

\title{Zero bias field selective transition adressing of the NV center via pulse shaping}

\author{
 Thomas Richard\\
  Département de génie électrique et de génie informatique\\Faculté de génie\\Université de Sherbrooke\\Sherbrooke, Québec, J1K 2R1, Canada\\\\
  Institut quantique\\ Université de Sherbrooke\\ Sherbrooke, Québec, J1K 2R1, Canada\\\\
  \texttt{thomas.richard2@usherbrooke.ca} \\
   \And
 Yves Bérubé-Lauzière \\
  Département de génie électrique et de génie informatique\\
  Faculté de génie\\
  Université de Sherbrooke,\\ Sherbrooke, Québec, J1K 2R1, Canada \\\\
  Institut quantique\\ Université de Sherbrooke\\ Sherbrooke, Québec, J1K 2R1, Canada\\\\
  \texttt{Yves.Berube-Lauziere@usherbrooke.ca}
}

\begin{document}
\maketitle
\begin{abstract}
Nitrogen–vacancy (NV) centers in diamond are a leading platform for vector magnetometry, offering intrinsic sensitivity to both the magnitude and direction of magnetic fields. NV magnetometers typically rely on an external bias field to lift the degeneracy of the ground-state spin sublevels, enabling spectral discrimination of different NV orientations and spin transitions. At high accuracy, however, such a bias field systematically introduces errors through thermal, mechanical, and hysteretic drifts, thus hindering the accurate measurement of the magnetic field of interest. Several strategies have been proposed to address this limitation, including optical polarization-based orientation labeling, optical anisotropy, strong coupling to nearby nuclear spins, circular microwave polarization, and tailored pulse sequences. In this work, we introduce a optimization--based pulse-shaping control framework that enables selective and robust manipulation of NV ensembles without relying on a static bias field. Our method achieves both orientation-selective and subspace-selective control through optimal temporal modulation of the driving fields, providing a route to resolving spectral overlaps in degenerate NV systems.

\end{abstract}


\section{\label{sec:level1}Introduction}

Nitrogen–vacancy (NV) centers in diamond have emerged as a powerful platform for magnetic field sensing, with applications in bioscience~\cite{su2025fluorescent}, solid-state physics~\cite{xu2023recent}, nanoscale
nuclear-magnetic-resonance spectroscopy~\cite{lovchinsky2016nuclear}, to name a few. Owing to their crystallographic site symmetry in the diamond lattice, NV centers provide intrinsic vectorial sensitivity, making them attractive candidates for accurate vector magnetometry\cite{taylor2008high}. 

In most implementations, NV magnetometers rely on an external bias field in order to clearly separate from each other the magnetic-field-dependent resonance frequencies of the NV center, enabling first-order sensitivity to magnetic fields and reconstruction of the magnetic field vector. Such a bias field can, however, be undesirable in many situations.
At the~nT measurement accuracy level, drifts in the bias field, caused by thermal~\cite{danieli2013highly}, mechanical, or hysteresis-induced drifts~\cite{moree2023review}, may be mistaken for drifts in the external field to be measured, leading to inaccuracies.
Moreover, in applications such as zero- to ultralow-field (ZULF) NMR spectroscopy~\cite{blanchard2007zero}, the study of ferromagnetic thin films~\cite{zazvorka2020skyrmion}, or experiments conducted in magnetically shielded environments, the bias field can significantly perturb or even disrupt the system under investigation. Consequently, bias-field-free operation has become an increasingly important objective, motivating the development of alternative vector reconstruction schemes.

At zero bias field, there is an overlap in the spectral features, preventing magnetometry based on frequency-resolved transitions. Nevertheless, the different NV orientations can be distinguished by distinct microwave driving rates (when microwaves are polarized along a sufficiently low-symmetry direction), or by their different optical emission patterns and polarizations. Moreover, selection rules permit addressing transitions between a chosen subspace of the spin 1 sublevels, \textit{e.g.} $m_s = 0$ to $m_s = 1$, making it possible to retain magnetic field sensitivity even at zero field. Hence, a method enabling selective discrimination and addressing of individual NV orientations or spin transitions is required to perform vector magnetometry without a bias field.

For orientation-selective magnetometry, several groups have exploited optical anisotropy to associate NV centers with their spatial emission patterns\cite{chen2020calibration, backlund2017diamond, weggler2020determination}, or used excitation and emission polarization to identify the corresponding NV orientation\cite{li2024vector, li2024simultaneous}. As for subspace control, methods based on strong coupling to nearby $^{13}$C\cite{wang2022zero} and circular microwave polarization\cite{lenz2021magnetic, zheng2019zero, alegre2007polarization} have also been proposed. These last methods were proven useful on a single NV, and on ensembles. However, methods developed on ensembles still rely on the optical polarization dependence of the emission to distinguish NV orientations. Finally, a method for obtaining full selectivity based on pulse shaping was also proposed for a spin locking experiment\cite{liddy2023optimal}.

\begin{figure}
    \centering
\includegraphics[width=0.5\linewidth]{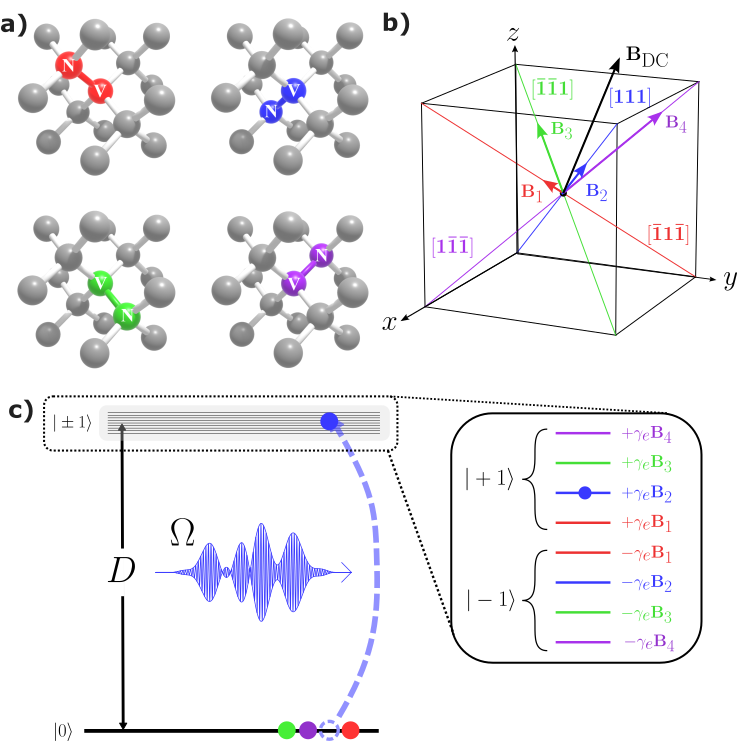}
    \caption{(a) The four possible crystallographic orientations of the NV center in the diamond lattice. The colored spheres represent nitrogen-vacancy centers, while the gray spheres represent carbon atoms. (b) Projection of an external magnetic field onto the quantization axis of each of the four NV orientations. (c) Selective spin manipulation achieved through optimal pulse shaping. In this example, the $|0\rangle \rightarrow |+1\rangle$ transition of the $[111]$-oriented NV center (blue) is selectively driven without the application of a bias magnetic field, while the remaining NV centers stay in the $|0\rangle$ ground state. The inset shows the corresponding energy-level structure induced by the external magnetic field.}

    \label{fig: Qualitative_picture}
\end{figure}

In the present work, the goal is to synthesize a target gate that selectively polarizes the spin associated with only one NV orientation and for one energy transition without affecting the other orientations or transitions, while being realizable experimentally. This manipulation is depicted in Fig.~\ref{fig: Qualitative_picture}~(c).
For this purpose, the aforementioned approaches are extended  by introducing an optimization--based pulse-shaping control framework that enables both subspace-selective and orientation-selective manipulation of NV center ensembles without the need for a bias magnet, providing a route toward optimized ensemble spin control. This framework establishes the feasibility of selective spin control in NV ensembles and opens opportunities for new applications (\textit{e.g.} ensemble Ramsey magnetometry). In the following, the theoretical framework and simulation results are presented, illustrating the performance of the approach proposed herein, and demonstrating that selectivity can be maintained even in the near-zero magnetic field regime and with perturbing electric fields.

This paper is organized as follows. In Sect.~\ref{sect: OptCrtlProb}, the optimal control problem underlying pulse shaping is formulated. Then, NV center system dynamics are described in Sect.~\ref{sect:NVSystDyn}, which introduces the single-NV model, the control-driven dynamics, and their extension to ensemble behavior. Controllability aspects are then discussed to assess the extent to which the system can be steered under the available controls. In Sect.~\ref{sect: CtrlObj}, the control objectives are defined through appropriate figures of merit. Sect.~\ref{sect: EnsSelTrans} presents the application of the framework to both an ideal system and more realistic scenarios, along with robustess aspects highlighting the impact of imperfections on the optimized control strategies. Finally, Sect.~\ref{sect: Conclusion} concludes the paper.
 
\section{Optimal Control Problem}\label{sect: OptCrtlProb}
Without the ability to precisely manipulate quantum systems, exploring their properties and leveraging them for quantum technologies is nearly impossible. Quantum optimal control (QOC)  theory~\cite{glaser2015training} addresses this challenge by optimizing the shape of dynamical controls to achieve desired objectives with high precision QOC has been extensively investigated for NV centers, with applications ranging from quantum sensing, to quantum computation and quantum information processing\cite{rembold2020introduction}. 

The system dynamics are governed by the unitary time-evolution operator $\hat U(t)$, according to the equation
\begin{equation}
|\psi(t)\rangle = \hat U(t)\,|\psi(0)\rangle,
\end{equation}
with $\hat U(t)$ obeying the Schr\"odinger equation
\begin{equation}
i\hbar\,\dot{\hat U}(t) = \hat H(t)\,\hat U(t),
\end{equation}
where $\hat H(t)$ is the system Hamiltonian. In typical practical situations, the Hamiltonian can be decomposed into a drift term and a set of externally controlled terms as follows
\begin{equation}\label{eq: bilinear form}
\hat H(t) = \hat H^{d} + \sum_j \Omega_j(t)\,\hat H^{c}_j,
\end{equation}
where $\Omega_j(t)$ are time-dependent control inputs (or fields) to the system. This yields the standard bilinear control system equation
\begin{equation}
\dot{ U}(t) =-\frac{i}{\hbar}\left(\hat H^{d} + \sum_j \Omega_j(t)\,\hat H^{c}_j\right)\hat U(t).
\end{equation}
The goal of QOC is to determine the set of control inputs $\Omega_j(t)$ to steer a quantum system from an initial state $|\psi(0)\rangle$ to a desired target state $|\psi_G\rangle$. Within this framework, numerical optimal control algorithms—most notably Gradient Ascent Pulse Engineering (GRAPE) have been developed to compute control fields that maximize a chosen figure of merit, such as state-transfer fidelity or gate fidelity\cite{khaneja2005optimal}.

The following introduces the essential components of the QOC formalism specialized to the control of NV center ensembles.
\section{NV center system dynamics}\label{sect:NVSystDyn}
\subsection{Single NV}\label{section:System Dynamics}

The NV center is a point defect of $C_{3v}$ symmetry in the diamond lattice consisting of a substitutional nitrogen atom  adjacent to a vacancy. The center's electronic structure consists of a  $^3A_2$ ground triplet state, an optically addressable $^3E$ triplet excited state and several singlet dark states. One of the key properties of an NV center lies in the ability to optically initialize its spin state towards the spin projection $m_s = 0$ via optical pumping, typically using a 532~nm green laser.
Microwaves (MW) can then be used to drive transitions from this ground state energy to the magnetically sensitive spin states $m_s  = \pm 1$. Choosing the $\hat z$ axis along the NV principal axis, the NV center ground state energy structure is well described by the Hamiltonian~\cite{doherty2012theory} 
\begin{equation}\label{eq:Hamiltonian}
    \hat{H}_{\rm NV}/\hbar =(D + \mathcal{E}_z)\left(\hat S_z^2 - \tfrac{2}{3}\right) + g_e\mu_B\mathbf{B}_{\rm DC}\cdot\hat{\mathbf{S}}+\mathcal{E}_x(\hat S_y^2 - \hat S_x^2) +  \mathcal{E}_y(\hat S_x \hat S_y + \hat S_y \hat S_x)+ \hat{\mathbf{S}}\cdot \mathbf{A}\cdot \hat{\mathbf{I}}, 
\end{equation}
where $\hbar$ is the reduced Planck constant, $g_e$ is the electron Landé factor, $\mu_B$ is the Bohr magneton, $\mathbf{B}_{\rm DC}$ is the ambient magnetic field vector, $\mathbf{\hat S} = [\hat S_x, \hat S_y, \hat S_z]^T$ is the spin-1 vector operator, $D / (2\pi) \approx2.87$~GHz is the axial zero field splitting, $\hat{\mathbf{I}} = [\hat I_x. \hat I_y, \hat I_z]^T$ is the nuclear spin vector operator of the neighboing $^{14}$N nucleus, and $\mathbf{A} = \operatorname{diag}(A_\perp, A_\perp,A_\parallel)$ is the diagonal hyperfine tensor with $A_\perp/(2\pi) = 2.7$~MHz and $A_\parallel/(2\pi) = 2.16$~MHz. The combined effect of strain and local electric field are described by $\mathcal{E}_{x,y} = d_\perp^E E_{x,y} + d_\perp^\sigma\sigma_{x,y}$ and $\mathcal{E}_z = d_\parallel^E E_{z} + d_\parallel^\sigma\sigma_{z}$, with couplings originating from the dipolar moments $d_\perp$ and $d_\parallel$\cite{doherty2013nitrogen}. Without loss of generality, $\mathcal{E}_z \approx 0$ will be assumed for it can be seen as an overall offset of the $D$ parameter. For the remainder of this paper, $\hbar =1$ will be set for simplicity. 

By applying the unitary transformation $\hat{\mathcal{U}}(t) = e^{-i\omega \hat S_z^2 t}$ and the rotating wave approximation, the resulting transformed drift Hamiltonian is given by 
\begin{equation}\label{eq:RWA Hamiltonian}
    \tilde{H}^d
    = \delta\left(\hat S_z^2 - \tfrac{2}{3}\right)  + (\mathcal{B}_z + m_IA_\parallel) \hat{S}_z 
+ \mathcal{E}_x (\hat{S}_y^2 - \hat{S}_x^2) + \mathcal{E}_y (\hat{S}_x \hat{S}_y + \hat{S}_y \hat{S}_x),
\end{equation}
where $\delta = D - \omega$ and $\mathcal{B}_z = g_e\mu_BB_z$. Here, non-secular terms have been ignored by way of the rotating wave approximation and a fixed nuclear spin number $m_I$ is considered (see appendix~\ref{appendix:RWA}). As a result, the magnetic interaction is fully characterized by the axial projection of the ambient magnetic field onto the NV axis (Fig.~\ref{fig: Qualitative_picture}(b)).

\subsection{Control dynamics}
The control Hamiltonians is generated by the projection of a MW source on the NV's principal axis system (PAS). The effective control Hamiltonian as seen by an NV center is given by  
\begin{equation}
    \hat H^c(t) = g_e \mu_B B(t) \cos(\nu t + \varphi(t)) \hat{\mathbf{u}} \cdot \hat{\mathbf{S}}
    = \epsilon(t)\cos(\nu t + \varphi(t))\mathbf{\hat u}\cdot \mathbf{\hat S},
\end{equation}
where $\nu$ is the carrier frequency, $\epsilon(t)= g_e\mu_BB(t)$ and $\varphi(t)$ are a tunable field and phase respectively, and $\hat{\mathbf{u}}$ is the unit vector in the direction of the microwave magnetic field. Expressing this microwave source in the rotating frame defined previously with $\nu = D = \omega$, the transformed Hamiltonian can be expressed as 
\begin{equation}\label{eq:RWA control Hamiltonian}
\tilde H^c(t) \overset{\rm RWA}{\approx}
\tfrac{1}{2}\epsilon(t)\hat{\mathbf u}\!\cdot\!
\begin{pmatrix}
\hat S_x\cos\varphi(t)-\hat S_x'\sin\varphi(t)\\
\hat S_y\cos\varphi(t)-\hat S_y'\sin\varphi(t)\\
0
\end{pmatrix},
\end{equation}
where $\hat S_{x,y}' = i[\hat S_z^2, \hat S_{x,y}]$ are transformed spin operators. The bilinear structure of Eq.~\eqref{eq: bilinear form} can be obtained by  direct factoring of the control Hamiltonian as shown in appendix~\ref{appendix: bilinear form}, and is given by
\begin{equation}
\tilde H^c(t) = \underbrace{\begin{bmatrix}
    \Omega_1(t) & \Omega_2(t)
\end{bmatrix}}_{\boldsymbol{\Omega}(t)} \underbrace{\begin{bmatrix}
    \mathbf{\hat u}_\perp&0\\0&-\mathbf{\hat u}_\perp
\end{bmatrix}\begin{bmatrix}
    \hat S_x\\\hat S_y\\ \hat S_x'\\\hat S_y'\\
\end{bmatrix}}_\mathbf{U\cdot \tilde S  = H^{c}},
\end{equation}
where ${\Omega_1(t) = \epsilon(t)\sin\varphi(t)}$ and ${\Omega_2(t) = \epsilon(t)\cos\varphi(t)}$, and $\mathbf{\hat u}_\perp$ represents the transverse component of the projection of the source along the $\mathbf{\hat e}_\perp$ axis of the NV PAS. The inclusion of additional control sources follows directly from the bilinear structure. Each source introduces new control amplitudes, which are concatenated into the vector $\boldsymbol{\Omega}(t) = [\Omega_1(t), \Omega_2(t)]
$, along with corresponding rows in the matrix $\mathbf{U}$ encoding their projections. For the purpose of optimal control, $\Omega_1(t)$ and $\Omega_2(t)$ can be considered as functionals of $\varepsilon(t)$ and $\varphi(t)$ over which the control will be optimized (see appendix~\ref{appendix: bilinear form} below).
 
To achieve full controllability, the dynamical Lie algebra (DLA) must satisfy the Lie algebra rank condition (LARC)~\cite{d2021introduction}. This condition dictates that a system is controllable if its DLA spans the entire state space, meaning  $\dim (\mathfrak{g}) = \dim (\mathfrak{su(n)})  = n^2-1 = 8 $. This condition can be achieved using a pair of independent, ideally orthogonal, control sources. It is worth noting that controllability can still be attained with non-orthogonal configurations, albeit at the cost of a higher conditioning (see subsection below).

\subsection{NV ensembles}

The NV center can be positioned along one of four tetrahedrally oriented crystallographic axes. While the Hamiltonian representation is straightforward for a single NV center, extending the description to an ensemble requires accounting for the distinct spatial orientation of each defect, as the projection of the control field depends on the NV axis.

Let \(N_s\) denote the total number of NV centers in the ensemble, each indexed with $s$ and oriented along one of the four tetrahedral crystallographic axes indexed by \(o \in \{1,2,3,4\}\), corresponding to the \(\langle 111\rangle\) symmetry directions denoted NV$_1$, NV$_2$, NV$_3$, and NV$_4$. These orientations are represented by the set of unit vectors
\begin{align*}
    &\rm NV_1: \tfrac{1}{\sqrt{3}}[1,1,1]&&
    \rm NV_2: \tfrac{1}{\sqrt{3}}[-1,-1,1],\\
    &\rm NV_3: \tfrac{1}{\sqrt{3}}[-1,1,-1]&& 
    \rm NV_4: \tfrac{1}{\sqrt{3}}[1,-1,-1],
\end{align*}
expressed in the crystal frame (see Fig.~\ref{fig: Qualitative_picture}~(a)). Each orientation is associated with a rotation matrix $\mathcal{R}_{\text{NV}_{o} \leftarrow \text{lab}}$ that maps vectors from the laboratory frame to the corresponding NV PAS~\cite{liddy2023optimal}. An ensemble is therefore fully specified by an assignment map
\[
\sigma : \{1,\dots,N_s\} \to \{1,2,3,4\},
\]
which associates to the NV center with index $s$ an orientation $\sigma(s)$. The number of centers in each orientation is denoted $N_o$, such that $\sum_{o=1}^{4} N_o = N_s$. In the case of an ideally uniform distribution, one has $N_o = \frac{N_s}{4},\forall o$, assuming $N_s$ is divisible by $4$ (\textit{e.g.}, $N_s = 48$ yields $N_o = 12$ centers per orientation).

Since no interaction between distinct NV centers is assumed, each subsystem evolves independently. Consequently, operators associated with different NV centers only act on their respective Hilbert spaces $\mathcal{H}^{(s)}$ and obey
\[
[{\hat S_k}^{(s)}, {\hat S_l}^{(s')}] = 0, \quad \forall\, s \neq s',\, k,l \in \{x,y,z\}, 
\]
where the label in parenthesis denotes the associated NV space. The DLA generated by the total Hamiltonian therefore decomposes as a direct sum
\[
\mathfrak{g}_{\rm ensemble} \cong \bigoplus_{s=1}^{N_s} \mathfrak{su}(3).
\]
Using the fundamental representation of each $\mathfrak{su}(3)$ subalgebra, the full system admits a block-diagonal matrix representation composed of $3\times 3$ blocks, yielding an overall dimension of $3N_s \times 3N_s$. This avoids the exponential scaling of the full tensor-product space ($3^{N_s}$), and instead provides a representation that scales linearly with the number of NV subsystems.

With this notation, the ensemble Hamiltonian can be written as a direct sum over all subsystems
\begin{equation}
\tilde{H}(t) = ~\bigoplus_{s=1}^{N_s} \left( \tilde H^{d, (s)} + \sum_{j=1}^{2N_c} \Omega_j(t)\, \tilde H^{c
,(s)}_j \right),
\end{equation}
where 
\begin{equation}\label{eq: Operator rotation}
\tilde H^{c, (s)}_j = \mathcal{R}_{\sigma(s)\leftarrow\rm lab}\mathbf{\hat u}\cdot \mathbf{\tilde S}^{(s)}
\end{equation}
denotes the control Hamiltonian acting on the NV center belonging to the subspace $s$. 

Within the bilinear formalism, the factorized form can be constructed by concatenation of all subsystems such that 
\begin{align*}
    \boldsymbol{\Omega}^{\rm ensemble} &= 
    \begin{bmatrix}
        \Omega_1(t) & \cdots & \Omega_{2N_c}(t)
    \end{bmatrix},\\[6pt]
    \mathbf{U}^{\rm ensemble} &= 
    \begin{bmatrix}
        \mathbf{U}^{(1,1)} & \cdots & \mathbf{U}^{(1,N_s)} \\
        \vdots & \ddots & \vdots \\
        \mathbf{U}^{(2N_c,1)} & \cdots & \mathbf{U}^{(2N_c,N_s)}
    \end{bmatrix},\\[6pt]
    \mathbf{\tilde S}^{\rm ensemble} &= 
    \begin{bmatrix}
        \mathbf{\tilde S}^{(1)} &
        \mathbf{\tilde S}^{(2)} &
        \dots &
        \mathbf{\tilde S}^{(N_s)}
    \end{bmatrix}^T,
\end{align*}
where $\mathbf{U}^{(j,s)}$ encodes the geometric projection of source $j$ onto the PAS of NV center $s$.

\subsection{Controllability of the ensemble}\label{sect: CtrlEnsemble}
To ensure controllability, the accessible control Hamiltonians must generate a DLA that spans the entire tangent space, which is here the operator space. This DLA is obtained recursively by evaluating repeated commutators until closure is reached. However, controllability only establishes that these directions span the space; it does not quantify how efficiently they do so.

To characterize efficiency, we evaluate the condition number of the generator basis (see Appendix~\ref{appendix: Angle between sources} for more details). A well-conditioned basis consists of nearly orthogonal generators, indicating that accessible directions are well-distributed throughout the Lie algebra. Conversely, a poorly conditioned basis contains nearly linearly dependent generators. This implies that certain directions can only be synthesized through delicate cancellations between large control components, reducing robustness.

For numerical evaluation, each generator (operator matrix) is flattened into a column vector. For \(\mathfrak{su}(n)\), these traceless anti-Hermitian matrices span an ($n^2$-1)-dimensional real vector space. The generator matrix is constructed by stacking these vectors side by side
\begin{equation}
M(\mathfrak{g}) = \begin{bmatrix} \mathrm{vec}(G_{1}) & \mathrm{vec}(G_{2}) & \cdots & \mathrm{vec}(G_{d}) \end{bmatrix}, \quad G_i \in \mathfrak{g}.
\end{equation}
The numerical conditioning is quantified by the condition number 
\begin{equation}
    \kappa (M)=\frac{\sigma _{\max }(M)}{\sigma _{\min }(M)},
\end{equation}
where $\sigma _{\max }$ and $\sigma _{\min }$ are the extreme singular values. A value of $\kappa(M)\approx 1$ signals an almost orthogonal basis, while a large value indicates near linear dependence. A concrete example is provided in Appendix~\ref{appendix: Angle between sources}.

Since the control Hamiltonians $\tilde H^{c,(s)}_j$ are determined by Eq.~\eqref{eq: Operator rotation}, the resulting ensemble DLA, defined as
\begin{equation} 
\mathfrak{g}_\mathrm{E}(\mathbf{\hat u}_1,\mathbf{\hat u}_2) = \left\{ \tilde H^{d,(s)}, \tilde H^{c,(s)}_j \,\middle|\, j=1,\dots,4,\; s=1,\dots,N_s \right\}_{\rm Lie}, 
\end{equation} 
where $\{\cdot \}_{\rm Lie}$ denotes the smallest Lie algebra generated by the specified set (initial set), \textit{i.e.} the linear span of all elements obtained through iterated commutator (the Lie closure of the set). The representation of the DLA depends on the source orientations through the control Hamiltonians. Consequently, the generator matrix $M$ and its condition number $\kappa$ inherits the dependences on  $\mathbf{\hat u}_1$ and $\mathbf{\hat u}_2$, allowing the numerical properties of the system to be directly optimized through the source configuration.

The optimal orientations can be obtained by minimizing the objective function
\begin{equation}\label{eq: PSO Cost function}
J(\mathbf{\hat{u}}_1, \mathbf{\hat{u}}_2)
=
\alpha
\log\!\left(
\kappa\!\left(
M(\mathfrak{g}_\mathrm{E}(\mathbf{\hat{u}}_1,\mathbf{\hat{u}}_2))
\right)
\right)
-
\beta
\frac{
\dim\!\left(
\mathfrak{g}_{\mathrm{E}}(\mathbf{\hat{u}}_1,\mathbf{\hat{u}}_2)
\right)
}{
N_s(n^2-1)
},
\end{equation}
where $\alpha$ and $\beta$ are positive weighting coefficients that balance the relative importance of numerical conditioning and controllability. A logarithmic scaling on the first term is employed to account for the exponential behavior of the condition number, preventing large values of $\kappa$ from dominating the optimization prooblem. The second term quantifies the degree of controllability by normalizing the dimension of the effective Lie algebra by its theoretical maximum, $N_s(n^2-1)$, thereby restricting the metric to the interval $[0,1]$. Since greater controllability is desirable, this term is included with a negative sign so that maximizing the Lie algebra dimension reduces the overall cost.

\begin{figure}[h]
\centering
\includegraphics[width=0.5\linewidth]{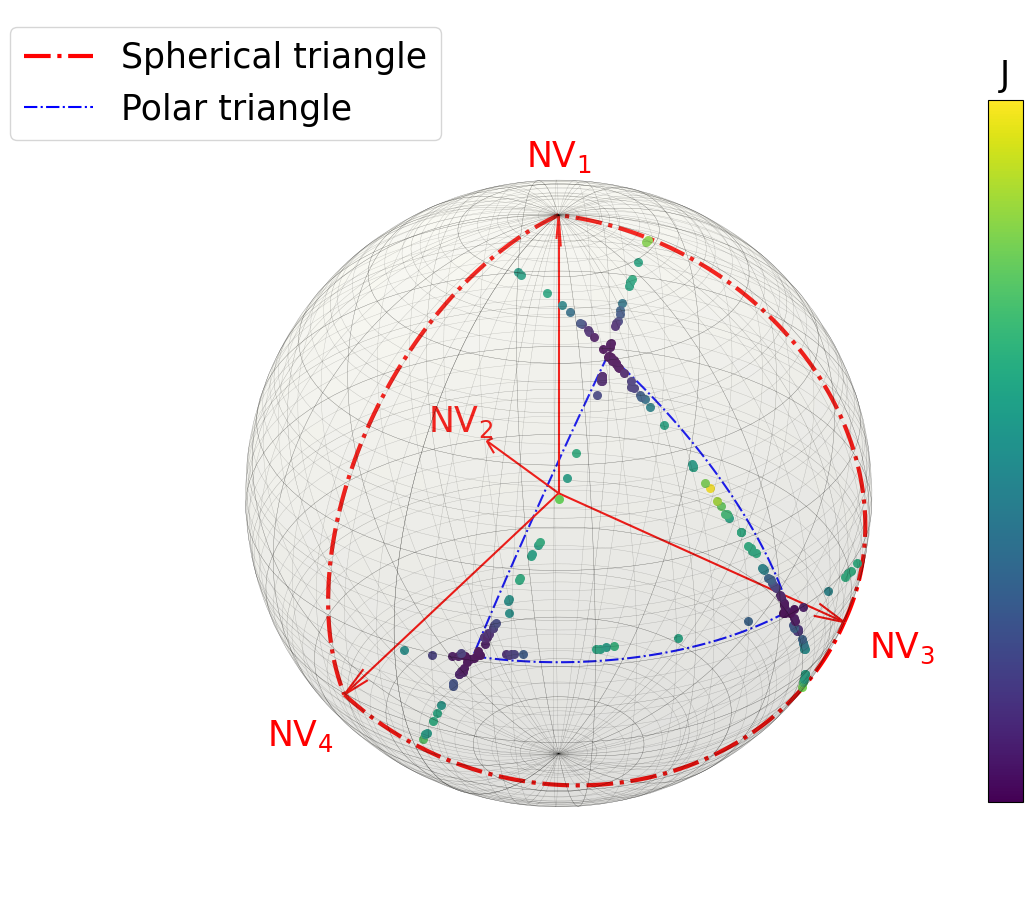}
     \caption{Optimal configuration of the first source obtained by PSO. The optimal ($\theta_1^\ast, \phi_1^\ast$) coordinates align along one of the polar triangles (blue) whose vertices are associated with the spherical triangle formed by the $\rm NV_{1,3, 4}$ orientations (red). The color scale indicates low values (blue) to high values (yellow) of the objective function $J$.}
    \label{fig:OrientationPSO}
\end{figure}

To perform the optimization, the source orientations $\mathbf{\hat u}_j$, $j=1,2$ were parametrized by using their polar and azimuthal angles, $\theta_j$ and $\phi_j$, defined in the laboratory frame. Because $\kappa$ lacks a closed-form and smoothly differentiable expression with respect to these control parameters, gradient-based optimization methods are not suited for this task. Instead, a particle swarm optimization (PSO) algorithm was used to search over the parameter space $(\theta_1, \phi_1, \theta_2, \phi_2)$. Exploiting the $C_{3v}$ symmetry of the NV centers, the search space for the first source was restricted to the domain $\theta_1, \phi_1 \in [0, 2\pi/3]$. For visualization purposes, the resulting optimal configurations for the first source $\mathbf{\hat u}_1$ appear in Fig.~\ref{fig:OrientationPSO}.

\begin{figure}
     \centering
    \includegraphics[width=0.5\linewidth]{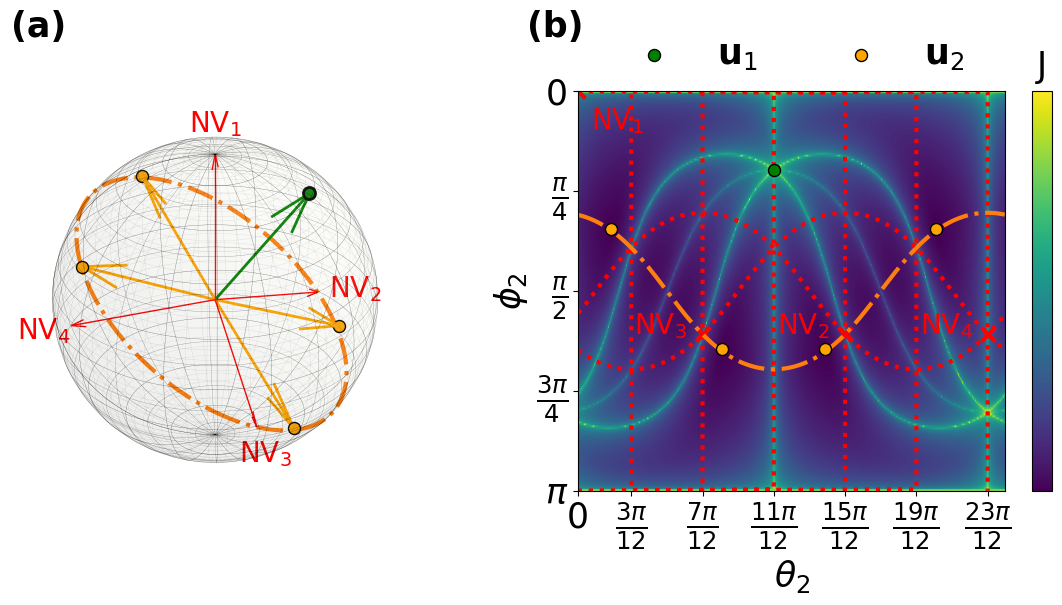}
     \caption{Optimal orientation for $\mathbf{\hat u}_2$ (orange arrows and orange dots) with $\mathbf{\hat u}_1$ (green arrow and green dot) fixed. (a) The orange arrows lie in the plane spanned by the orientations of the $\mathrm{NV}_2$ and $\mathrm{NV}_3$ centers. (b) Orthographic projection of the value of the cost function for $\mathbf{\hat u_2}$. Red crosses ("$\times$") correspond to the NV orientations and red dotted lines to the plane spanned by NV orientation pairs. The orange dashed line represents the plane spanned by the optimal configurations for the second source. As expected, planes spanned by the first source and an NV axis result in bad conditioning (high values in the color scale) due to their collinearity.}
    \label{fig:OrientationPSO2}
 \end{figure}

The extrema occur at the vertices of the polar triangle (blue triangle in Fig.~\ref{fig:OrientationPSO}) associated with the spherical triangle defined by three NV orientations (red triangle in Fig.~\ref{fig:OrientationPSO})~\cite{todhunter1863spherical}. 
Fixing the first source orientation $(\theta_1, \phi_1)$ to one of these optimal extrema reduces the optimization space, allowing a direct visualization of the cost function with respect to the remaining parameters of the second source. The resulting two-dimensional profile over $(\theta_2, \phi_2)$ appears in Fig.~\ref{fig:OrientationPSO2}.

Choosing $\hat{\mathbf u}_1$ to be coplanar with the plane spanned by $\mathrm{NV}_1$ and $\mathrm{NV}_2$, it is found that the optimal orientation of $\hat{\mathbf u}_2$ lies in the plane spanned by $\mathrm{NV}_3$ and $\mathrm{NV}_4$. This configuration consistently yields higher control fidelities than the conventional orthogonal basis.

\section{Control objective}\label{sect: CtrlObj}

As mentioned in the introduction, the control objective is to synthesize a target gate that selectively adresses angular specific spin transitions associated with only one NV orientation, while being experimentally feasible. In practice, the implemented (optimized) gate is denoted $\hat{U}_F$, which approximates the targeted unitary gate $\hat{U}_G$. The targeted gate is therefore defined such that the desired operation $\hat{U}_G$ is applied to all NV centers aligned along the selected orientation, whereas the identity operator $\hat{\mathbb{I}}$ is applied to all NV centers corresponding to the other orientations. A gate targeting the $\rm NV$ $i^\mathrm{th}$ orientation can thus be constructed as 
\begin{equation}
    \hat{ U}_G = \bigoplus_{s=1}^{N_s}\hat U_G^{(s)},\quad \text{where} \quad \hat U_G^{(s)} =\begin{cases}
        \hat U_G&\text{if }\sigma(s) = \rm NV_{i}\\
        \mathbb{\hat I}&\text{otherwise}.
    \end{cases}
\end{equation}

In this work, we focus on subspace-selective gates, as defined in Appendix~\ref{appendix: pseudo_spin_1/2}, which effectively implement a pseudo-spin-$\tfrac{1}{2}$ operation within a restricted subspace of the full spin-1 Hilbert space. To this end, we introduce a weighted cost function of the form
\begin{equation}
    \Phi = \Phi_{\rm infidelity} + \lambda_{\rm BW}\Phi_{\rm bandwidth} + \lambda_{\rm BC}\Phi_{\rm boundary},
\end{equation}
where
\begin{align*}
    \Phi_{\rm infidelity} & = 1 - \frac{1}{d^2}\left|\operatorname{tr}\left(\hat U_G^\dagger \hat U_F\right)\right|^2, \\ 
    \Phi_{\rm bandwidth} &= \sum_j \frac{1}{T}\int_0^T |\dot \Omega_j(t)|^2 dt,\\
    \Phi_{\rm boundary} &= \sum_j \bigl(\Omega_j(0)^2 + \Omega_j(T)^2\bigr).
\end{align*}
Here $\Phi_{\rm infidelity}$ quantifies the overlap between the simulated $\hat U_F$ and targeted gate $\hat U_G$,  $\Phi_{\rm bandwidth}$ penalizes rapid variations of the control amplitudes, and $\Phi_{\rm boundary}$ enforces zero values at the beginning and end of the pulse. 

The optimization is performed using the GRAPE algorithm. In the standard GRAPE discretization, each control field is expressed as a series of piecewise-constant control amplitudes $u_j(t_k)\in \mathcal{U}\subseteq \mathbb R$ with $t_k \in [0,T]$, where $T$ is the total duration and $\mathcal{U}$ denotes the subset of admissible control amplitudes. The total duration $T$ is divided into $M$ time intervals of equal width $\Delta t = t_k - t_{k-1} = T/M$ with $k=1,2,\dots,M$ . The propagator associated with the time interval $[t_{k-1},t_k)$ is then given by
\begin{equation}
\hat U_k = 
\exp\!\left\{-\frac{i\Delta t}{\hbar}
\left(\hat H^d +  \boldsymbol{\Omega}(t_k)\mathbf{\hat H}^c\right)\right\}.
\end{equation}
Setting the initial and final values $\hat U_0 = \hat{\mathbb{I}}$ and $\hat U_{M+1} = \hat U_G$ , the infidelity cost function becomes
\begin{equation}
\Phi_{\rm infidelity}
=
1 - \frac{1}{d^2}
\left|
\operatorname{tr}
\left(
\hat U_{M+1}
\hat U_M
\hat U_{M-1}
\cdots
\hat U_0
\right)
\right|^2 .
\end{equation}

Applying the chain rule, a component of the gradient of the total cost function with respect to the control amplitude $\Omega_j(t_k)$ is given by
\begin{align*}
\tfrac{\partial \Phi}{\partial \Omega_j(t_k)} &=
\tfrac{1}{N}\,
\mathrm{Re}\!\left\{
\operatorname{tr}\!
\left(
e^{-i\phi_g}
\hat \Lambda_{M+1:k+1}^{\dagger}
\frac{\partial \hat U_k}{\partial \Omega_j(t_k)}
\hat X_{k-1:0}
\right)
\right\} \\[4pt]
&\quad +
\tfrac{\lambda_{\rm BW}\Delta t}{M}
(2\mathbf{\mathsf{D}}^T\mathbf{\mathsf{D}}\Omega_j)_k \\[4pt]
&\quad +
2\lambda_{\rm BC}\!\left(
\Omega_j(0)\delta[t_k]
+
\Omega_j(T)\delta[t_k-T]
\right),
\end{align*}
where $\hat X_{k:0} := \hat U_k \cdots \hat U_1 \hat U_0$ is the forward-propagated gate, $\hat \Lambda^\dagger_{M+1:k+1} := \hat U_G \hat U_M \hat U_{M-1} \cdots \hat U_{k+1}$ is the backward-propagated (adjoint) gate, and  $\delta[t] = 1$ if $t = 0$ and 0 otherwise. The phase factor is defined as
\[
\phi_g =
\arg\!\left(
\tfrac{1}{N}
\operatorname{tr}\!\left(
\hat U_G^\dagger \hat U_F
\right)
\right),
\]
and $\mathsf D$ denotes a finite-difference operator used to approximate the time derivative in the bandwidth penalty (appendix~\ref{Appendix: Difference_Operator}). 

Using this gradient, the control amplitudes are iteratively updated according to
\begin{equation}
\Omega_j^{(l+1)}(t_k)
=
\Omega_j^{(l)}(t_k)
-
\varepsilon
\frac{\partial \Phi}{\partial \Omega_j^{(l)}(t_k)},
\end{equation}
where $\varepsilon$ is a step size determined through a standard line-search procedure (\textit{e.g.}, L-BFGS).
\section{Ensemble selective transition}\label{sect: EnsSelTrans}

\subsection{Ideal system}

\begin{figure*}
\centering
\includegraphics[width=1\linewidth]{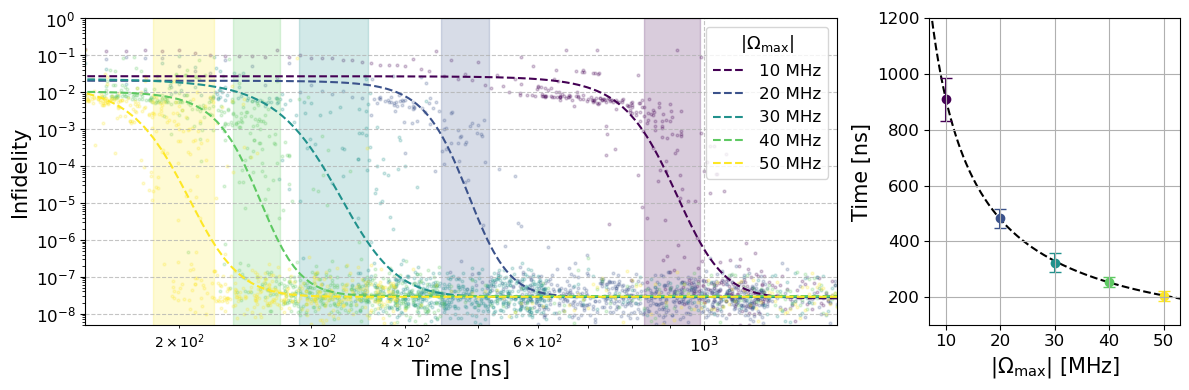}
\caption{Left: Infidelity as a function of pulse duration for amplitude bounds (see values in inset for $|\Omega_{\rm max}|$). The data were fitted to the logistic model $\log(\Phi_{\rm infidelity}(T)) = L\left(1+e^{-k(T-T_{\rm QSL}^\ast)}\right)^{-1}$, where $T^\ast$ provides an estimate of the minimal control time. The associated $[25\%,75\%]$ confidence intervals are also shown as the colored shaded vertical bars. Right: Estimated minimal times as a function of maximal pulse amplitude. Optimal times in increasing order of pulse magnitude are given by $T^\ast_{\rm QSL} = [909\pm80, 480\pm 40, 320\pm 40, 250\pm 20,200\pm 20]$~ns. The black dotted line corresponds to an inverse power-law fit of the estimated minimal times.}
\label{fig:minimal_time}
\end{figure*}

In order to design the control pulses, the total duration $T$ must be specified. In the absence of prior knowledge of the intrinsic quantum speed limit $T_{\rm QSL}$, an estimate of the minimal achievable duration, denoted $T^\ast$, is obtained empirically. Because the optimal duration depends on the maximum allowed control amplitude, $T^\ast$ is extracted from numerical optimizations performed under varying amplitude constraints, using a minimal model of four NV centers corresponding to the crystallographic orientations. $T^\ast$ is defined as the midpoint of the convergence transition in the infidelity (see Fig.~\ref{fig:minimal_time}, left).

The results indicate that the optimal pulse duration scales approximately inversely with the maximum control amplitude (Fig.~\ref{fig:minimal_time} - right). Furthermore, near the minimal time, the control amplitudes exhibit strong saturation effects, frequently reaching the imposed amplitude bounds (not shown). Additionally, in these conditions, the controls displayed rapid and large fluctuations between high and low amplitude values, resulting in highly non-smooth pulse profiles. Such behavior is likely to be experimentally unfeasible, as it requires both operation at hardware limits and fast temporal modulation beyond realistic bandwidth constraints. This can be attributed to the optimization procedure, which prioritizes fidelity without enforcing smoothness or regularity of the control fields. To mitigate these effects and improve both numerical stability and experimental feasibility, slightly longer pulse durations are therefore preferable in practice and were used subsequently.

\begin{figure*}
\includegraphics[width = \linewidth]{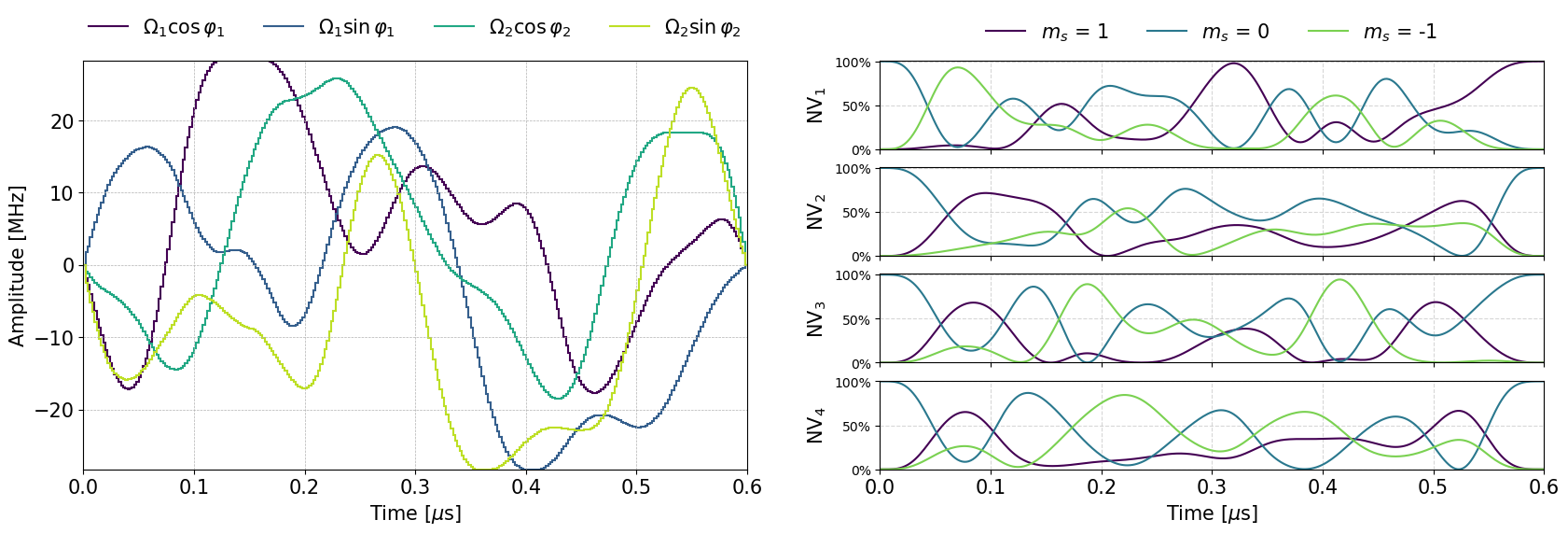}
\caption{Simulation result for the ensemble optimization with $\mu_Bg_e\mathbf{B}_{\rm ext} = [-0.25,-1.62,0.26]$~MHz, $T = 600$~ns and $\Delta t = 2$~ns. Left: Control amplitudes during the implementation of the gate $\hat U_G=\exp(-i\pi\hat S_x^+)$ on the $[111]$ orientation. The $z$ axis of the laboratory frame is defined along the $z$ axis of the diamond cristal. Right: Population dynamics for each NV orientations. A final fidelity of 99.99998\% for $m_s = 1$ for NV$_1$ and $m_s = 0$ for NV$_2$, NV$_3$, and NV$_4$ was obtained for this implementation}
\label{fig:Single NV evolution}
\end{figure*}

Using a maximum control amplitude
\begin{equation}
|\Omega_{\max}| = \max_{t \in [0,T]} \sqrt{\Omega_1^2(t) + \Omega_2^2(t)} = 40~\text{MHz},
\end{equation}
the NV ensemble was simulated in the presence of a weak external magnetic field of magnitude $\|\mathbf{B}_{\rm ext}\| = 60$~~\textmu T, while electric perturbations were neglected. The total pulse duration was chosen slightly above the estimated speed limit, with $T = 600~\text{ns}$ and a time step $\Delta t = 2~\text{ns}$. The control objective was to implement a $\pi$-pulse on the $\mathrm{NV}_1$ orientation within the $\{0,+1\}$ subspace, around its $\hat{\mathbf{e}}_x$ axis, that is
\begin{equation}
\hat U_G = \exp\!\left(-i\pi \hat S_x^{+}\right).
\end{equation}
The resulting dynamics are shown in Fig.~\ref{fig:Single NV evolution}.
Using this control strategy, high fidelities were consistently obtained regardless of the initial pulse shape, suggesting the absence of problematic local minima in the optimization landscape. Sinusoidal initial pulses nevertheless produced smoother final control profiles compared to other trial shapes.

 
\subsection{Robust excitation}\label{sect:Robust excitation}

To implement selective control pulses within a realistic magnetometry protocol, the optimization must be robust against both unknown external magnetic fields and inhomogeneities in the internal electric field parameters. To include these effects, deviations to the drift Hamiltonian are added as follows 
\begin{equation}
\tilde{H}(t) = \bigoplus_{s=1}^{N_s} \left( \tilde H^{d, (s)}(\mathbf{b}_s, \mathbf{e}_s) + \sum_{j=1}^{2N_c} \Omega_j(t)\, \tilde H^{c,(s)}_j \right),
\end{equation}
where $\mathbf{b}_s$ and $\mathbf{e}_s$ denote magnetic and electric field vectors. These fields are randomly drawn from prescribed distributions to emulate ensemble inhomogeneities. For low-strain ensembles, the electric field magnitude is typically on the order of $5\times 10^{5}$~$\tfrac{\rm V}{\rm m}\sim 100$~kHz, while the magnetic field can vary within the interval $[-75,75]$~\textmu T corresponding to approximately $[-2,2]$~MHz in frequency units.

\noindent In this simulation, the targeted unitary for was choosen to be
\[
\hat U_G =
\exp\!\left\{-i\frac{\pi}{2}\hat S_x^{+}\right\}
\]
applied solely to the $\rm NV_1$ orientations among the ensemble.

For the optimization procedure, $N_s = 50$ NV centers were used. The magnetic field amplitudes were drawn uniformly within a spherical volume with $\|\mathbf{b}_i\| \in [0,2]~\text{MHz}$, reflecting the absence of assumptions on the local field orientation or strength. The electric field amplitudes $\|\mathbf{e}_i\|$ were sampled from a Gamma distribution $\Gamma(k=3,\theta=0.025)$, leading to a most probable value near $100~\text{kHz}$. This distribution is well suited due to its positive support and flexible skewed shape, with low weight near zero and a decaying tail at higher amplitudes. Here, it is not assumed to be exact, but used as a phenomenological choice to model and probe unknown ensemble inhomogeneities.

\begin{figure}
    \centering
    \includegraphics[width=0.5\linewidth]{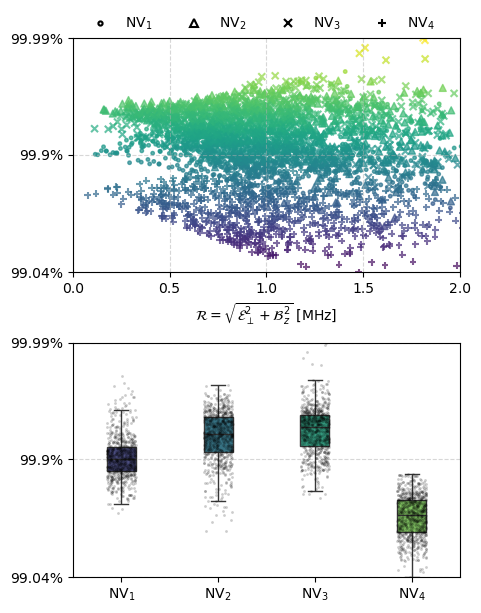}
    \caption{Robust implementation of a selective $\hat U_G = \exp\{-i\pi\hat S_x^+\}$ pulse for the $[111]$ orientation. Top: Distribution of the fidelity as a function of perturbation amplitudes. Bottom:  Fidelity distribution as a function of NV orientation. Central boxes represent the interquartile range (IQR) while the internal horizontal line denotes the median. Whiskers extend to $1.5 \times \text{IQR}$. }
    \label{fig:Ramsey}
\end{figure}

To validate pulse optimization, the solution was evaluated on a larger Monte Carlo test ensemble of 3000 NV centers, with $\|\mathbf{b}_i\|$ and $\|\mathbf{e}_i\|$ uniformly distributed in the intervals $[0,2]~\text{MHz}$ and $[0,1]~\text{MHz}$, respectively in order to capture electric field distribution dependence of the optimization procedure. The results shown in Fig.~\ref{fig:Ramsey} indicate transition fidelities exceeding $99.0\%$ for all simulated NV centers.

Only a weak dependence on the direction of the perturbing fields was observed. In contrast, the fidelity exhibits a field-dependent variance (heteroscedasticity) with respect to the field amplitudes across many simulated systems. Also, no significant variation in pulse fidelity was found when modifying the electric field distribution used in the optimizer. Additionally, a dependence of the control efficiency on the NV orientation can be observed.  Although the results for the NV$_2$ and NV$_3$ orientations displayed in Fig.~\ref{fig:Ramsey} yield the highest state-transfer fidelities, this behavior was not consistently observed across multiple solutions obtained with the same settings. These predicted high fidelities now call for experimental validation.

\section{Conclusion}\label{sect: Conclusion}
In this work, we developed a comprehensive quantum optimal control framework for selective state transition manipulation of NV center ensembles, explicitly accounting for NV crystallographic orientation, control field projections, and experimentally relevant constraints. By formulating the ensemble dynamics as a block-diagonal representation in the fundamental irreducible representation of $\mathfrak{su}(3)$, we reduced the effective system dimension while preserving full controllability of each subspace. This approach enables scalable optimization without resorting to the exponentially large tensor-product Hilbert space.

We first demonstrated the feasibility of orientation-selective and transition-selective control in an idealized four-orientation NV ensemble. An empirical analysis of the minimal pulse duration revealed a clear inverse scaling with the maximum control amplitude, as well as strong saturation effects near the optimal time. These observations motivated the use of slightly longer pulse durations to ensure numerical stability and experimental robustness. As a result, high-fidelity selective gates were obtained consistently, largely independent of the pulse initial guess.

We then extended the model for a robust ensemble control setting by incorporating random magnetic and electric field inhomogeneities. Using a Monte Carlo ensemble within the GRAPE optimization, we synthesized control pulses that maintain high selectivity and achieve high fidelities around $99\%$ over thousands of randomly sampled NV configurations. The results indicate that selective excitation remains effective across a broad range of field amplitudes and orientations, although heteroscedastic behavior with respect to field strength was observed. This highlights the importance of explicitly including disorder and uncertainty in the control design.

Overall, this study establishes a scalable and physically consistent methodology for ensemble-selective quantum control in NV systems. The framework provides a practical route towards vector magnetometry and orientation-resolved sensing protocols. Future work will focus on experimental validation, incorporation of decoherence and pulse-shaping bandwidth constraints, and the exploration of analytical control solutions suggested by the observed symmetry in the dynamics.

\appendix
\section{Rotating-frame approximation for the NV center}
\label{appendix:RWA}

To suppress the rapidly oscillating contribution associated with the zero-field splitting (ZFS) in Eq.~\eqref{eq:Hamiltonian}, the unitary transformation
\begin{equation}\label{eq: transformation unitaire}
\hat{\mathcal{U}}=e^{i\omega \hat S_z^2 t}
\end{equation}
is used. Since the operators $\hat S_z$, $\hat S_x \hat S_y+\hat S_y \hat S_x$, and $\hat S_x^2-\hat S_y^2$ commute with $\hat S_z^2$, only the transverse spin components $\hat S_x$ and $\hat S_y$ are affected by this transformation. Using
\[
e^{-i\omega \hat S_z^2 t}\hat S_{x,y}e^{i\omega \hat S_z^2 t}
=\cos(\omega t)\hat S_{x,y}+\sin(\omega t)\hat S'_{x,y},
\]
where $\hat S'_{x,y} = i[\hat S_z^2,\hat S_{x,y}]$, the Hamiltonian in the rotating frame becomes
\begin{equation}
\begin{aligned}
\hat H_{\rm NV} &=
\left(D+\mathcal{E}_z-\omega\right)\!\left(\hat S_z^2-\tfrac{2}{3}\right) \\
&\quad +\mathcal{E}_x(\hat S_y^2-\hat S_x^2)
+ \mathcal{E}_y(\hat S_x\hat S_y+\hat S_y\hat S_x) \\
&\quad + \left(A_\perp \hat I_x + g_e\mu_B B_x\right)
\!\left[\hat S_x\cos(\omega t)+\hat S'_x\sin(\omega t)\right] \\
&\quad + \left(A_\perp \hat I_y + g_e\mu_B B_y\right)
\!\left[\hat S_y\cos(\omega t)+\hat S'_y\sin(\omega t)\right] \\
&\quad + (A_\parallel \hat I_z + g_e\mu_B B_z)\hat S_z.
\end{aligned}
\end{equation}

Here, the nuclear spin projection $m_I$ is treated as a quasi-static parameter, taking values in $\{-1,0,1\}$, since the $^{14}\mathrm{N}$ nuclear spin dynamics are strongly suppressed by the quadrupolar interaction on the timescale of the control pulses. Consequently, the axial hyperfine term acts as an effective static contribution proportional to $m_I$, while the quadrupolar shift can be absorbed into the effective detuning.

Choosing the unitary transformation frequency as 
\[
\omega = D_{\rm eff} = D + \mathcal{E}_z,
\]
and neglecting all rapidly oscillating terms by the rotating wave approximation (RWA), the effective Hamiltonian reduces to
\begin{equation}
\hat H_{\rm NV} \overset{\rm RWA}{\approx}
\mathcal{E}_x(\hat S_y^2-\hat S_x^2)
+ \mathcal{E}_y(\hat S_x\hat S_y+\hat S_y\hat S_x)
+ (m_I A_\parallel + g_e\mu_B B_z)\hat S_z .
\end{equation}

All contributions from the transverse magnetic field and transverse hyperfine coupling are suppressed in this frame due to the large ZFS. The remaining axial hyperfine interaction can therefore be interpreted as an effective static bias field along the NV axis given by
\[
B_{\rm eff} = \frac{A_\parallel m_I}{g_e\mu_B}.
\]

For the control field, transforming to the rotating frame gives
\begin{equation}
\tilde H^c(t)
= \epsilon(t)\cos(\nu t+\varphi(t))\,
\hat{\mathbf u}\!\cdot\!
\begin{pmatrix}
\hat S_x\cos(\omega t)+\hat S_x'\sin(\omega t)\\
\hat S_y\cos(\omega t)+\hat S_y'\sin(\omega t)\\
\hat S_z
\end{pmatrix}.
\end{equation}
Using standard trigonometric identities and choosing the resonant condition $\nu = \omega = D_{\rm eff}$, the counter-rotating contributions oscillating at $2\omega$ can be neglected. The control Hamiltonian thus reduces, under the RWA, to
the form given in Eq.~\eqref{eq:RWA control Hamiltonian} with a matrix representation given by 
\begin{equation}
    \tilde H^c(t) = \epsilon(t)u_\perp\begin{pmatrix}
        0&e^{-i(\theta_u - \varphi(t))}&0\\
        e^{i(\theta_u - \varphi(t))}&0&e^{-i(\theta_u + \varphi(t))}\\0&e^{i(\theta_u + \varphi(t))}&0
    \end{pmatrix},
\end{equation}
where $u_\perp = \sqrt{u_x^2 + u_y^2}$ and $\tan\theta_u = u_y/u_x$.

It is to be noted that the RWA follows from time-dependent perturbation theory where fast oscillating terms contribute at first order as $\mathcal{O}(\epsilon/\omega)$, while their leading second-order correction produces a Bloch--Siegert shift of order $\epsilon^2/(4\omega)$. The approximation is therefore controlled by the dimensionless parameter
\[
\gamma = \frac{\epsilon}{2\omega}.
\]
Imposing the conservative condition $\gamma < 10^{-2}$ and using $\omega / 2\pi = 2.87~\mathrm{GHz}$ yields a breakdown threshold of
\[
\epsilon_{\max} \approx 60~\mathrm{MHz}.
\]
\section{Bilinear and vectorial formulation}\label{appendix: bilinear form}

The bilinear form of the control can be obtained by reorganizing the terms of  the control Hamiltonian. By factoring the spin operator terms in  Eq.~\eqref{eq:RWA control Hamiltonian}, the control Hamiltonian can be written as
\begin{equation}
\begin{aligned}
    \tilde H^{c} &=\epsilon(t)\cos\varphi(t) \left(\frac{u_x\hat S_x + u_y\hat S_y}{2}\right) - \epsilon(t)\sin\varphi(t) \left(\frac{u_x\hat S_x' + u_y\hat S_y'}{2}\right)\\
    &\equiv \Omega_1(t)\tilde H^{c}_1 + \Omega_2(t)\tilde H^{c}_2,
\end{aligned}
\end{equation}
where 
\begin{equation*}
    \Omega_1(\epsilon, \varphi) = \epsilon(t)\cos\varphi(t),\quad\Omega_2(\epsilon, \varphi) = \epsilon(t)\sin\varphi(t)
\end{equation*}
are functionnals of the control parameters and $u_{x,y}$ are the source components (in the NV reference frame). It is to be noted that the control of two degrees of freedom is available with a single source. 

The control Hamiltonian can be directly factorized as 
\begin{equation}
    \tilde H^{c} = \boldsymbol{\Omega}\cdot \mathbf{U}\cdot \mathbf{\tilde S} =  \boldsymbol{\Omega}\cdot \mathbf{\hat H}^{c},
\end{equation}
where 
\begin{align*}
    \boldsymbol{\Omega} &= \begin{bmatrix}
        \Omega_1(t)&\Omega_2(t)&\dots & \Omega_{2N_c}(t)
    \end{bmatrix},\\
    \mathbf{\tilde S}&=\begin{bmatrix}
        \hat S_x & \hat S_y & \hat S_x' & \hat S_y'
    \end{bmatrix}^T,\\
    \mathbf{U} &= \frac{1}{2}\begin{bmatrix}
        u_{1x}&u_{1y}&0&0\\0&0&-u_{1x}&-u_{1y}\\
        \vdots &\vdots &\vdots &\vdots\\
         u_{N_cx}&u_{N_cy}&0&0\\0&0&-u_{N_cx}&-u_{N_cy}
    \end{bmatrix}\\ 
    &= \frac{1}{2}\begin{bmatrix}
        \mathbf{\hat u}_{\perp, 1}&0\\0&-  \mathbf{\hat u}_{\perp, 2}\\
        \mathbf{\hat u}_{\perp, 1}&0\\0&-  \mathbf{\hat u}_{\perp, 2}\\
        \vdots &\vdots &\\
        \mathbf{\hat u}_{\perp, N_c}&0\\0&-  \mathbf{\hat u}_{\perp, N_c}
    \end{bmatrix},
\end{align*}
where $\mathbf{\hat u}_{\perp,j}$ are the perpendicular components of source $j$ in the NV reference frame.

\section{Pseudo spin-$\tfrac{1}{2}$ basis}\label{appendix: pseudo_spin_1/2}

To clarify the effect of applying a pair of perpendicular control fields, it is often convenient to express the dynamics in a pseudo spin-$\tfrac{1}{2}$ basis.  
Consider two controls respectively along the $\mathbf{\hat e}_x$ and $\mathbf{\hat e}_y$ axes of the NV center. In the rotating frame, the total control Hamiltonian can be written as
\begin{equation}
    \tilde H^c= \tfrac{1}{2}\left[\Omega_1(t)\, \hat S_x + \Omega_2(t) \, \hat S_x'+ \Omega_3(t) \, \hat S_y + \Omega_4 (t) \, \hat S_y'\right].
\end{equation}
with $\Omega_{2i-1} = \epsilon_i(t)\cos\varphi_i(t)$ and $\Omega_{2i} = \epsilon_i(t)\sin\varphi_i(t)$, $i\in \{1,2\}$. Introducing the pseudo spin-$\tfrac{1}{2}$ operators
\begin{align*}
    \hat S_x^{+} &= \tfrac{\hat S_x + \hat S_y'}{\sqrt{2}} =
    \begin{pmatrix}0 & 1 & 0\\ 1 & 0 & 0\\ 0 & 0 & 0\end{pmatrix}, & 
    \hat S_x^{-} &= \tfrac{\hat S_x - \hat S_y'}{\sqrt{2}} =
    \begin{pmatrix}0 & 0 & 0\\ 0 & 0 & 1\\ 0 & 1 & 0\end{pmatrix}, \\
    \hat S_y^{+} &= \tfrac{\hat S_y - \hat S_x'}{\sqrt{2}} =
    \begin{pmatrix}0 & -i & 0\\ i & 0 & 0\\ 0 & 0 & 0\end{pmatrix}, &
    \hat S_y^{-} &= \tfrac{\hat S_y + \hat S_x'}{\sqrt{2}} =
    \begin{pmatrix}0 & 0 & 0\\ 0 & 0 & -i\\ 0 & i & 0\end{pmatrix},
\end{align*}
the control Hamiltonian can be written as
\begin{align*}
    \tilde H^c 
    =& \frac{1}{2\sqrt{2}}\Big[
        (\Omega_1(t) + \Omega_4(t))\, \hat S_x^+ 
      +(\Omega_1(t) - \Omega_4(t))\, \hat S_x^- + (\Omega_3(t) - \Omega_1(t)) \, \hat S_y^+ 
      + (\Omega_3(t) + \Omega_1(t))\, \hat S_y^-
      \Big] \\
    \equiv& \tilde{\Omega}_1(t) \hat S_x^+ + \tilde{\Omega}_2(t) \hat S_x^- + \tilde{\Omega}_3(t) \hat S_y^+ + \tilde{\Omega}_4(t) \hat S_y^-,
\end{align*}
This form explicitly relates the control amplitudes and phases $(\Omega_{1,2}, \varphi_{1,2})$ to the selective transitions in the pseudo spin-$\tfrac{1}{2}$ subspaces. 

\section{Optimal angle between sources: Single NV case}\label{appendix: Angle between sources}
The implementation of an arbitrary Hamiltonian within a controllable quantum system can be formulated as a decomposition problem in the DLA. Let the initial set of experimentally available control generators be
\begin{equation}
    \mathcal{C}=\{G_1,G_2,\ldots,G_m\},
\end{equation}
which spans the initial control subspace. The dynamical Lie algebra generated by these operators is defined as
\begin{equation}
    \mathfrak{g}=\{{G_1, G_2, \dots, G_j}\}_{\rm Lie},
\end{equation}
where the closure operation includes all linear combinations and repeated commutators of the generators.

The generation of new Hamiltonian directions through commutation follows from the group commutator operation (the Baker-Campbell-Hausdorff relation)
\begin{equation}
e^{\epsilon A}e^{\epsilon B}e^{-\epsilon A}e^{-\epsilon B}
=
e^{\epsilon^2[A,B]+\mathcal{O}(\epsilon^3)},
\end{equation}
which demonstrates that sequences of evolutions generated by accessible Hamiltonians $A$ and $B$ can synthesize an effective evolution along the commutator direction $[A,B]$. By recursively computing higher-order commutators, the complete set of accessible Hamiltonian directions is obtained and forms the dynamical Lie algebra $\mathfrak{g}$.

Once a basis of the DLA has been constructed, any Hamiltonian belonging to the controllable subspace can be expressed as a linear combination of these basis elements,
\begin{equation}
    H=\sum_{i=1}^{n}x_i G_i ,
\end{equation}
where $x_i$ are the expansion coefficients. By vectorizing the operators in an orthonormal operator basis, this relation can be written as a linear system,
\begin{equation}\label{eq: H=Mx}
\hat H=M\mathbf{x},
\end{equation}
where $\mathbf{x}$ contains the expansion coefficients and 
\begin{equation*}
M=\begin{bmatrix}
\operatorname{vec}(G_1) & \operatorname{vec}(G_2) & \cdots & \operatorname{vec}(G_n)
\end{bmatrix}
\end{equation*}
is the matrix whose columns are the vectorized Lie algebra generators. The numerical robustness of the Hamiltonian decomposition is fundamentally governed by the conditioning of the generator matrix. To quantify this stability, consider a perturbation of the target Hamiltonian,
\begin{equation}
    \hat H \rightarrow \hat H + \delta\hat H,
\end{equation}
which consequently induces a variation in the expansion coefficients,
\begin{equation}
    \mathbf{x} \rightarrow \mathbf{x} + \delta\mathbf{x}.
\end{equation}
The resulting perturbed linear decomposition satisfies
\begin{equation}
    \mathrm{vec}(\hat H + \delta\hat H) = M(\mathbf{x} + \delta\mathbf{x}).
\end{equation}
Assuming that the perturbation remains strictly within the column space of $M$, the least-norm solution to this perturbed system is obtained via the Moore--Penrose pseudoinverse,
\begin{equation}
    \mathbf{x} + \delta\mathbf{x} = \tilde M \left( \mathrm{vec}(\hat H) + \mathrm{vec}(\delta\hat H) \right),
\end{equation}
where $\tilde M = (M^T M)^{-1}M^T$. Subtracting the unperturbed reconstruction relation isolates the perturbation of the reconstructed coefficients:
\begin{equation}
    \label{eq:error_mapping}
    \delta\mathbf{x} = \tilde M \mathrm{vec}(\delta\hat H).
\end{equation}

Taking the induced norm of Eq.~\eqref{eq:error_mapping} and invoking the submultiplicative property of matrix norms establishes the absolute error bound
\begin{equation}
    \|\delta\mathbf{x}\| \leq \|\tilde M\| \|\mathrm{vec}(\delta\hat H)\|.
\end{equation}
Concurrently, taking the norm of the forward relation $\mathrm{vec}(\hat H) = M\mathbf{x}$ yields $\|\mathrm{vec}(\hat H)\| \leq \|M\| \|\mathbf{x}\|$. Combining these two inequalities limits the relative reconstruction error to
\begin{equation}
    \label{eq:relative_error_bound}
    \frac{\|\delta\mathbf{x}\|} {\|\mathbf{x}\|} \leq \|M\| \|\tilde M\| \frac{ \|\mathrm{vec}(\delta\hat H)\| } { \|\mathrm{vec}(\hat H)\| }.
\end{equation}
Finally, by defining the generalized condition number of the rectangular matrix $M$ under the spectral norm as
\begin{equation}
    \kappa(M) \equiv \|M\| \|\tilde M\| = \frac{\sigma_\mathrm{max}(M)}{\sigma_{\mathrm{min}}(M)},
\end{equation}
the standard relative perturbation bound
\begin{equation}
    \frac{\|\delta\mathbf{x}\|} {\|\mathbf{x}\|} \leq \kappa(M) \frac{ \|\mathrm{vec}(\delta\hat H)\| } { \|\mathrm{vec}(\hat H)\| }
\end{equation}
is obtained. This expression provides a worst-case estimate of the sensitivity of the Hamiltonian reconstruction. The condition number $\kappa(M)$ quantifies the intrinsic amplification of this uncertainty due to the chosen generator basis. A well-conditioned generator matrix, characterized by a small value of $\kappa(M)$, ensures that experimental imperfections are
not significantly amplified during the reconstruction of the physical parameters. Conversely, an ill-conditioned basis leads to strong error amplification, making the inferred coefficients highly sensitive to small perturbations in the measured Hamiltonian.

To illustrate the proposed framework, a single NV center driven by two independent control sources is considered. The corresponding control fields are chosen as
\begin{equation}
    \mathbf{\hat u}_1=\Omega_1\mathbf{\hat x},
    \qquad
    \mathbf{\hat u}_2=\Omega_2\left(\cos\theta\,\mathbf{\hat x}
    +\sin\theta\,\mathbf{\hat y}\right),
\end{equation}
where $\theta$ denotes the angle between the two sources in the plane transverse to the NV axis. After transforming into the rotating frame, the initial set of generators is
\begin{align*}
   \mathcal{G}
   &=
   \left\{
   \hat S_x,\,
   \hat S_x',\,
   \cos\theta\,\hat S_x+\sin\theta\,\hat S_y,\,
   \cos\theta\,\hat S_x'+\sin\theta\,\hat S_y'
   \right\} \nonumber\\
   &=
   \{G_1,G_2,G_3(\theta),G_4(\theta)\}.
\end{align*}
The remaining generators are obtained by taking successive commutators,
\begin{align*}
    G_5 &= [G_1,G_2], &
    G_6 &= [G_1,G_3], \\
    G_7 &= [G_1,G_4], &
    G_8 &= [G_3,G_4],
\end{align*}
which close the dynamical Lie algebra under commutation. Vectorizing the eight generators produces the matrix

\begin{equation*}
\resizebox{\linewidth}{!}{$
    M=
    \begin{pmatrix}
\tfrac{1}{\sqrt{2}} & \tfrac{i}{\sqrt{2}} & \tfrac{1}{\sqrt{2}}e^{-i\theta} & \tfrac{i}{\sqrt{2}}e^{-i\theta} & 0 & 0 & 0 & 0 \\
\tfrac{1}{\sqrt{2}} & -\tfrac{i}{\sqrt{2}} & \tfrac{1}{\sqrt{2}}e^{i\theta} & -\tfrac{i}{\sqrt{2}}e^{i\theta} & 0 & 0 & 0 & 0 \\
\tfrac{1}{\sqrt{2}} & -\tfrac{i}{\sqrt{2}} & \tfrac{1}{\sqrt{2}}e^{-i\theta} & -\tfrac{i}{\sqrt{2}}e^{-i\theta} & 0 & 0 & 0 & 0 \\
\tfrac{1}{\sqrt{2}} & \tfrac{i}{\sqrt{2}} & \tfrac{1}{\sqrt{2}}e^{i\theta} & \tfrac{i}{\sqrt{2}}e^{i\theta} & 0 & 0 & 0 & 0 \\
0 & 0 & 0 & 0 & i\sin\theta & -i & -i\cos\theta & -i \\
0 & 0 & 0 & 0 & 0 & -i & -ie^{-i\theta} & -ie^{-2i\theta} \\
0 & 0 & 0 & 0 & 0 & 2i & 2i\cos\theta & 2i \\
0 & 0 & 0 & 0 & 0 & -i & -ie^{i\theta} & -ie^{2i\theta} \\
0 & 0 & 0 & 0 & -i\sin\theta & -i & -i\cos\theta & -i
\end{pmatrix},
$}
\end{equation*}
whose singular values are
\begin{align*}
\sigma_{1,2} &=2\cos\frac{\theta}{2},
&
\sigma_{3,4} &=2\sin\frac{\theta}{2},\\
\sigma_5 &=\sqrt{2}\sin\theta,
&
\sigma_6 &=2\sin\theta,
\end{align*}
\begin{align*}
\sigma_7
&=
\sqrt{
-5\sin^2\theta
+12
-\sqrt{\sin^4\theta-120\sin^2\theta+144}
},
\\
\sigma_8
&=
\sqrt{
-5\sin^2\theta
+12
+\sqrt{\sin^4\theta-120\sin^2\theta+144}
}.
\end{align*}
The condition number of the generator matrix is therefore completely determined by 
\begin{equation}
    \kappa(\theta)
    =
    \frac{\sigma_8(\theta)}
         {\min(\sigma_{1,2}(\theta),\sigma_{3,4}(\theta),\sigma_7(\theta))},
\end{equation}
 and reaches its minimum at $\theta=\frac{\pi}{2}$ corresponding to two mutually orthogonal control fields as expected. The same procedure extends naturally to the ensemble case, where the generators are defined according to Eq.~\eqref{eq: Operator rotation}.








\section{Finite difference operator}\label{Appendix: Difference_Operator}

To account for finite experimental bandwidth and associated instrumental constraints, the temporal regularity of the control fields is enforced through a quadratic penalty on their time derivatives. In discrete form, this requires the introduction of a finite-difference operator \(\mathsf{D}\), whose structure directly determines the spectral filtering properties of the regularization.

Any linear, translation-invariant discrete derivative operator can be written as
\begin{equation}
(\mathsf{D} \Omega)_n = \sum_{k \in \mathcal K} a_k \, \Omega_{n+k},
\end{equation}
where \(\mathcal K \subset \mathbb{Z}\) defines the stencil and \(a_k\) are the associated coefficients. Two standard choices are the forward difference operator, defined by \(\mathcal K_f = \{0,1\}\),
\begin{equation}
a_{f,0} = -\frac{1}{\Delta t}, \quad a_{f,1} = \frac{1}{\Delta t},
\end{equation}
and the centered difference operator, defined by \(\mathcal K_c = \{-1,0,1\}\),
\begin{equation}
a_{c,-1} = -\frac{1}{2\Delta t}, \quad a_{c,0} = 0, \quad a_{c,1} = \frac{1}{2\Delta t}.
\end{equation}

Due to translation invariance, these operators are diagonalized by discrete Fourier modes \(e^{i \omega n \Delta t}\), with associated symbols
\begin{equation}
\lambda(\omega) = \sum_{k \in \mathcal K} a_k e^{i \omega k \Delta t}.
\end{equation}
As a result, the quadratic penalty \(\|\mathsf{D}\Omega\|^2\) induces a frequency-dependent weighting proportional to \(|\lambda(\omega)|^2\), which determines how strongly each spectral component is suppressed.

For the forward and centered schemes, the corresponding spectral weights are
\begin{equation}
|\lambda_f(\omega)|^2 = \frac{1}{\Delta t^2}\sin^2\!\left(\frac{\omega \Delta t}{2}\right), \qquad
|\lambda_c(\omega)|^2 = \frac{1}{\Delta t^2}\sin^2(\omega \Delta t).
\end{equation}
These expressions illustrate a trade-off between accuracy and high-frequency suppression. The centered difference scheme is second-order accurate for derivatives but does not penalize the Nyquist frequency $\omega = \pi/\Delta t$, allowing grid-scale oscillations to persist. In contrast, the forward difference scheme suppresses this mode, but is only first-order accurate and introduces stronger numerical dispersion.

To combine the advantages of both schemes, we employ a mixed operator constructed as a weighted sum of their associated quadratic forms:
\begin{equation}
\|\mathsf{D} \Omega\|^2 \;\equiv\; \|\mathsf{D}_c \Omega\|^2 + \lambda_D^2 \|\mathsf{D}_f \Omega\|^2,
\end{equation}
where \(\mathsf{D}_c\) and \(\mathsf{D}_f\) denote the centered and forward difference operators, respectively, and \(\lambda_D\) is a tuning parameter.

This hybrid construction preserves the higher-order accuracy of the centered stencil for smooth components, while the forward term selectively suppresses spurious oscillations near the Nyquist frequency. In practice, a modest weighting (\(\lambda_D \sim 10^{-1}\)) is sufficient to regularize these high-frequency artifacts without significantly degrading the approximation quality at lower frequencies.

\paragraph{Funding.} This work was in part funded by the Quantum Sensing Program (QSP) of the Canada National Research Council (CNRC), Grant No. QSP-051-2.

\paragraph{Acknowledgments.} The authors thank L. Childress, R. Ruhlmann, and V. Halde for fruitful technical discussions.

\paragraph{Disclosures.} The authors declare
no conflicts of interest. 

\paragraph{Data availability.} Data underlying the results presented in this paper are
not publicly available at this time but may be obtained from the authors upon
reasonable request.

\nolinenumbers
\nocite{*}

\bibliographystyle{unsrt}  
\bibliography{references}  

\end{document}